# Partition Decomposition for Roll Call Data

G. Leibon[1,2], S. Pauls[2], D. N. Rockmore[2,3,4], and R. Savell[5]

Abstract

In this paper we bring to bear some new tools from statistical learning on the analysis of roll call data. We present a new data-driven model for roll call voting that is geometric in nature. We construct the model by adapting the "Partition Decoupling Method," an unsupervised learning technique originally developed for the analysis of families of time series, to produce a multiscale geometric description of a weighted network associated to a set of roll call votes. Central to this approach is the quantitative notion of a "motivation," a cluster-based and learned basis element that serves as a building block in the representation of roll call data. Motivations enable the formulation of a quantitative description of ideology and their data-dependent nature makes possible a quantitative analysis of the evolution of ideological factors. This approach is generally applicable to roll call data and we apply it in particular to the historical roll call voting of the U.S. House and Senate. This methodology provides a mechanism for estimating the dimension of the underlying action space. We determine that the dominant factors form a low- (one- or two-) dimensional representation with secondary factors adding higher-dimensional features. In this way our work supports and extends the findings of both Poole-Rosenthal and Heckman-Snyder concerning the dimensionality of the action space. We give a detailed analysis of several individual Senates and use the AdaBoost technique from statistical learning to determine those votes with the most powerful discriminatory value. When used as a predictive model, this geometric view significantly outperforms spatial models such as the Poole-Rosenthal DW-NOMINATE model and the Heckman-Snyder 6-factor model, both in raw accuracy as well as Aggregate Proportional Reduced Error (APRE).

[1] Memento, Inc., Burlington, MA, 01803
[2] Department of Mathematics, Dartmouth College, Hanover, NH 03755
[3] Department of Computer Science, Dartmouth College, Hanover, NH 03755
[4] The Santa Fe Institute, 1399 Hyde Park Road, Santa Fe, NM 87501
[5] Thayer School of Engineering, Dartmouth College, Hanover, NH 03755

# Introduction

Spatial models of parliamentary voting provide powerful geometric tools for the analysis of legislative bodies. Over the last three decades, spatial models have been developed and substantially refined. Important examples include the work of Cahoon (Cahoon 1975), Cahoon, Hinich and Ordeshook (Cahoon, et al., 1976), Hinich and Pollard (Hinich and Pollard 1981), Enelow and Hinich (Enlow and Hinich 1984), Hinich and Munger (Hinich and Munger 1994, 1997), Ordeshook (Ordeshook 1976, 1986), and Poole and Rosenthal (Poole and Rosenthal 1984, 1985, 2001, 2007). Spatial models produce a low-dimensional representation of the (presumed) very high-dimensional "action space," defined as the collection of "all contemporary political issues and government policies" (Ordeshook 1976, p. 308, as quoted in Poole 2005, p.14) that underlie the legislative process. These low-dimensional representations still hold an enormous amount of information, as measured by their ability to predict correctly the vast majority of votes cast.

The NOMINATE models of Poole and Rosenthal are of particular interest. When applied to the data of the aggregated roll call votes over all Congresses, they produce one- or two-dimensional spatial models that seem to explain the majority of voting behavior in the U.S. Congress (Poole and Rosenthal 2007). In their words, "voting is along ideological lines when positions are predictable across a wide set of issues" (Poole and Rosenthal 2007, p. 3). In this context "ideology" has a precise definition as their interpretation of the first dimension (the major dimension) of their spatial model, which is taken to be a measure of the extent to which a legislator supports government intervention in economic issues. The second dimension is broadly interpreted in terms of region, but in fact, the detailed Poole-Rosenthal analysis suggests a more complicated picture hiding beneath the surface. This is supported by the Heckman-Snyder work (Heckman and Snyder 2003) in which a factor model is used to show that an interaction of at least five spatial dimensions is necessary to explain fully roll call votes.

As a whole, these models (and others) show that the one-dimensional "liberal-conservative" axis of ideology, which appears to capture so much information, is in fact a dynamically changing complex amalgam of a variety of factors. In this paper, through a new form of analysis of roll call votes, we introduce a new geometric model for parliamentary voting that articulates the complex structure of ideology, providing a quantitative and data-driven formulation that allows us to study the nature and evolution of ideology as it is expressed in roll call votes. In particular, we adapt a new technique from statistical learning, the *Partition Decoupling Method* (PDM) (Leibon, et al. 2008), to the analysis of roll call data and apply it to historical roll call votes from the U.S. Congress. The PDM is a very general technique for the analysis of correlations in a family of high-dimensional feature vectors. It was originally applied to the analysis of time series of stock prices. In that context the PDM articulated the movement of stock prices as a linear combination of effects at various scales (e.g., market, sector, and industry) and revealed both the overall contribution and interaction of these effects. The main point of this paper is to show that an analogous (and informative) articulation of political structure can be accomplished by adapting these methods for the analysis of roll call data.

To adapt the PDM methodology to roll call data, a legislator's voting record is given a natural encoding as a point in a high-dimensional space: a history of an individual legislator's $m$ roll call votes (each encoded as either a 1, 0, or -1) is viewed as vector in a space of dimension $m$. This embedding of the voting records of a collection of legislators then sits inside an $m$-dimensional "roll call space." The distance between two legislators in roll call space is essentially a measure of the correlation of their roll call vote history. The identification of clusters in the data is the first step in constructing a data-driven definition of ideology and it provides a first form of dimension reduction: each cluster gives rise to a cluster-averaged roll call vote, which suitably normalized defines a cluster-based "motivation." The $m$ votes of each legislator are then summarized by a vector of length $N$ (where $N$ is the number of clusters, in practice much smaller than $m$) consisting of the weights measuring how close a legislator's votes are to each of the cluster-averaged roll call votes. When we remove this dimension-reduced data from the roll call votes, we are left with a residual data series for each legislator. We can then repeat the clustering process on the residuals to reveal a new, subsidiary dimension reduction, which can in turn be removed from the new data. This process is iterated as long as the residual data is distinguishable from random data.

The result is a model in which every legislator votes according to a set of weighted motivations defined in terms of the degree of a priori support (or the lack thereof) for each vote, and a legislator's voting record is a weighted sum of these motivations. Aggregating all the weights with respect to a given collection of motivations in a cluster of legislators determines a data driven notion of "ideology," effectively defined as a collection of positions with high average weight. We emphasize that these "ideological dimensions" are determined via unsupervised learning – that is, the number and description of the dimensions are determined by the data.

This model has a number of satisfying features. First, it produces a quantifiable and more textured description of ideology, making possible the discovery of those issues and policy positions dictated by ideological concerns. Second, the data-driven description of ideology provides a means to quantify its evolution. Third, it creates a simple framework in which positions that might seem incongruous (at least according to some conventional labeling) can be present simultaneously within a single ideology. This is in agreement with the seminal work of Converse (Converse 1964) that showed that, in general, individual ideological constructs are riddled with inconsistencies and contradictions. Note that this point of view is another feature that distinguishes our approach from spatial models: a main underlying assumption in the construction of spatial models is that legislators maintain consistent ideologies, while our method does not require such an assumption and, indeed, may be used to evaluate the validity of such a claim. Fourth, our model allows for the emergence of motivations for legislators with different strengths – weights for a given motivation may be positive, negative, large, small, or even zero. This feature is absent from spatial models, in which each legislator has an opinion positioned on each of the different axes (as indicated by their point in space) and these opinions are counted equally when compared to a cut associated to a vote.

We use our approach to analyze the roll call voting history of the House and Senate of each of the U.S. Congresses.[6] The first layer of the resulting multiscale view confirms the spatial model findings in that it exposes a dominant effect of party identification (see e.g., Poole and Rosenthal 2007). However, our PDM-inspired approach provides a more textured (yet still low-dimensional) description that goes beyond the one or two dimensions of the NOMINATE model or the multidimensional models of Heckman and Snyder. Our analysis decomposes each major ideology into sub-ideologies and the accompanying geometric analysis provides descriptive axes for each legislative body and coordinates for each legislator. Successive layers produce closer and closer approximations to the exact roll call data and ultimately represent the data as a linear combination of ideological motivations for each legislator. From this we obtain predictive models for voting that we can compare to spatial models via correct classification percentage and APRE. In both cases, our methods significantly outperform both the NOMINATE model and the Heckman-Snyder 6-factor model.

Our exposition of the methodology is augmented by the presentation of detailed PDM-based analyses of the 108$^{th}$, 88$^{th}$, and 77$^{th}$ Senates. These examples provide a good demonstration of the variation in behaviors capable of being detected by the PDM. In the case of the 108$^{th}$ Senate, we see that the first layer is essentially one-dimensional and party-based while the second layer (appearing on a finer scale) reflects a new partition of the legislators distinct from the party-based partition of the first layer. In the 77$^{th}$ and 88$^{th}$ Senates the initial layer of clusters (and hence the legislators) cannot be arranged along a single dimension. In the 88$^{th}$ Senate, we again find a second layer dominated by issue-based concerns rather than party, while in the 77$^{th}$ the residual data exhibits no additional detectable structure. Our issue-based characterization of the second layer of the 108$^{th}$ and 88$^{th}$ Senates is similar to the factors identified by Poole-Rosenthal (Poole and Rosenthal 2007) as aspects of the second dimension of their spatial model. Our analysis differs from spatial models in that our explanation of the second layer uses a completely different geometric model. In a second novel use of statistical learning we identify the "best" separating issues via the statistical approach of AdaBoost (Freund and Schapire 1997, 1999). This yields a qualitative description of issues that split the larger party ideologies into partially conflicting sub-ideologies. Our procedure also gives us a quantitative comparison between the two layers, allowing us to measure their relative strength. In contrast, the methods used to understand the dimensions produced in spatial models are based on necessarily subjective classifications of roll call votes (see e.g., Poole and Rosenthal 2007).

---

[6] We used the data available at Keith Poole's "Voteview Website", http://www.voteview.com.

In summary, our approach and the tools we use are novel in three ways. First, we use unsupervised methods to locate geometrically significant clusters in the network associated to the roll call data. Second, the iterative method of removing and reexamining the residual data for additional structure is entirely new. Third, our use of the AdaBoost algorithm to help describe qualitatively the motivations is novel in this context and is quantitative. We expect this approach to be of great use in future analyses of roll call data, either independently or as a companion to spatial models.

## Spatial models

We briefly review some aspects of the spatial models in the literature: the models of Poole and Rosenthal (Poole and Rosenthal 2007), the Bayesian estimation of ideal points of Clinton, Jackman and Rivers (Clinton, et al. 2004), and the linear probability model of Heckman and Snyder (Heckman and Snyder 2004).

Consider the simplest one-dimensional spatial model in the context of a simple example of a single bill voted on by a collection of legislators. In this case, each bill and vote is assigned an *ideal point* on the real line. The line is meant to encode the position of the bill relative to some characteristic. For each bill a legislator examines its ideal point with respect to her own position on the bill relative to that ideal point: if the vote position is to the left of the bill, she votes "yes" and if to the right, she votes "no." Higher-dimensional models are constructed analogously: bills are assigned an ideal hyperplane in $n$-dimensional Euclidean space. Votes are cast according to the spatial relationship of the ideal points and the bill: a legislator will vote "yes" if to one side of the hyperplane and "no" if to the other.

In their work over the last two decades, Poole and Rosenthal have shown that the parsimonious model described above holds great power. If the ideal positions are calculated using a maximum likelihood estimator from the roll call data, the resulting voting model correctly predicts over 80% of all votes in the history of the U.S. Congress. Moreover, they find, with the exception of two periods in history, that a one- or two- dimensional model is sufficient to explain the vast majority of votes. Poole, Rosenthal and their co-authors have refined these ideas over the last three decades, producing the family of NOMINATE scoring systems based on these spatial models (cf., Poole and Rosenthal 1984, 1985, 2007; McCarty, et al. 1997, 2006; Poole 2005).

Clinton, Jackman and Rivers (Clinton, et al. 2004) establish a Bayesian approach for evaluating legislative preferences from roll call data in the context of one or more models of legislative behavior. Their framework is broadly applicable and can be adapted easily to focus on issues of specific interest or to incorporate additional aspects or assumptions. As an application, their methods can be used to estimate ideal points for spatial models of voting. Their work allows for a much more refined (and faster) probabilistic estimation of ideal points with very few restrictions on the underlying data. Moreover, their techniques, unlike those of Poole-Rosenthal, allow for an accurate analysis of error in the ideal point estimations.

Heckman and Snyder (Heckman and Snyder 2003) present another approach. They construct a linear probability model to which a factor analysis is applied. Their results show that roll call voting is not determined by just one or two dimensions, but often many more (usually at least five), indicating a more complex structure. The Heckman-Snyder model has a computational advantage – it is linear and hence computationally inexpensive. Its output is the collection of significant eigendata[7] associated to a similarity matrix derived from the roll call data. If, a priori, the number of significant eigenvectors is fixed and they are used as coordinates of ideal points in a Euclidean space, they recover a close analogue[8] of the Poole-Rosenthal spatial model.

## The Partition Decoupling Method (PDM)

Key to our methodology is an adaptation of the Partition Decoupling Method (PDM) of Leibon, et al. (Leibon, et al. 2008). Although originally developed for the analysis of a collection of time series data, the PDM provides a general approach to the analysis and articulation of multiscale characteristics in high-dimensional data. In particular, the PDM produces a multiscale (i.e., hierarchical) model for roll call voting. We explicitly point out that the resulting model is not a spatial model in the sense described above.

For a given legislative body (e.g., U.S. Senate or House of Representatives) over a fixed timeframe let $Leg(i)$ denote the voting record of legislator $i$ ($i = 1,…,n$). It will be a vector of length $m$, where $m$ is the number of votes in that session. The $j^{th}$ entry of $Leg(i)$ records the vote of legislator $i$ on bill $j$. It is equal to *1* or *-1*, depending on whether legislator *i* voted "yea" or "nay," and it is *0* in the case that legislator *i* either was not present, did not vote, or voted "present" on the *j*[th] bill. The *roll call matrix, V,* is the *n x m* array of votes. To use only relevant data, we remove all votes with a minority of less than 2.5% of the total body[9].

---

[7] "Eigendata" refers to the eigenvalues and eigenvectors of the matrix.

[8] This is particularly clear if we recall a theoretical result of Poole (Poole 2005) who shows that if voting is perfect (according to the spatial model) then the first nontrivial eigenvector of the double-centered roll call matrix is exactly the ideal point values for the one-dimensional spatial model. The eigendata of the double-centered matrix is essentially what Heckman and Snyder use in their analysis.

Our model represents each legislator's roll call record as a weighted sum of basic "motivations" $M_j$,

$$Leg(i) = \sum_{j=1}^{N} \alpha_j(i) M_j. \quad (1.1)$$

A motivation, $M_j$, is a unit vector of length *m*. The entries of the motivation $M_j$ encode the strength of support (or opposition in the case of a negative weight) associated to each of the individual votes. In this way the motivation vector is the record of an ideal voter with respect to the issues underlying the motivation. For example, if $M_j(k)$ is positive and large, we interpret this as indicating that the ideal voter associated with the motivation would vote "yes" on the *k*[th] vote. Similarly, if $M_j(k)$ is large and negative, the corresponding ideal voter would vote "no" on the *k*[th] vote. The weights $\alpha_j(i)$ then represent the individual weights for legislator *i* that reflect how tied each legislator is to a given motivation, i.e., how closely a legislator's true voting matches the ideal voting encoded in the motivation vector.

The flexibility in choice and number *N* of the motivations (all we require is unit length and the ability to represent the roll call data) is at the heart of a more subtle definition of ideology than a black and white yea-nay decision on a given vote. Motivations can either be declared a priori or discovered from the geometry of the roll call space via clustering. We pursue the latter approach, extracting plausible candidates for the motivations (and the corresponding weights) according to where the majority of the information in the data lies. Fort this, the basic object of study is the *n x n correlation matrix,*[10] *S* determined by the roll call matrix. The entries in *S* are defined by

$$S_{i,j} = corr(Leg(i), Leg(j)) = \frac{1}{N-1} \frac{Leg(i)- <Leg(i)>}{std(Leg(i))} \cdot \frac{Leg(j)- <Leg(i)>}{std(Leg(j))}. \tag{2.1}$$

---

[9] We choose this threshold to make our analysis comparable to that of Poole and Rosenthal (Poole and Rosenthal 2007).

Note that the correlation between legislators *i* and *j* is simply the cosine of the angle between the centered legislator vote vectors in *m*-dimensional space and is easily related to the distance between the points defining the legislators in the *m*-dimensional roll call space. The matrix *S* also can be thought of as defining a *weighted network* of legislators in which the strength of connection between any two legislators is given by the correlation of their roll call vectors.

From this geometric embedding of the legislators we derive the motivations via a partitioning of the legislators into geometrically determined clusters. Each cluster determines a motivation given by the normalized mean (centroid) of the cluster. Generally, these motivations will be linearly independent. The first "layer" in the roll call data is then given by its projection onto these motivations. The residual relative to these motivations is obtained by subtracting this first projection. The process of clustering, constructing motivations, and projecting, and finally, computing residuals, can now be repeated until the residual is indistinguishable from noise.

We'll now be more precise: let $k_0$ denote the number of clusters found in the original normalized roll call data. For each cluster we create an average voting profile by taking the (normalized) mean of all votes over the members of the cluster. Let $\{M_1^1, \ldots, M_{k_0}^1\}$ denote these first scale (or level) motivations. To calculate the weights $\alpha_j(i)$ we project the vote vector $Leg(i)$ of the $i^{th}$ legislator onto the subspace spanned by $\{M_1^1, \ldots, M_{k_0}^1\}$. The weights $\{\alpha_j^1(i)\}_{j=1}^{k_0}$ are the coefficients of the projected vector written in terms of the motivations $\{M_1^1, \ldots, M_{k_0}^1\}$.[11] From this we derive a first "scale" approximation to the roll call space, given by

$$A^{(1)} = A^{(1)}_{\{M_1,\ldots M_{k_0}\}} = \begin{pmatrix} \sum_{j=1}^{k_0} \alpha_j^1(1) M_j^1 \\ \vdots \\ \sum_{j=1}^{k_0} \alpha_j^1(n) M_j^1 \end{pmatrix}$$

---

[10] There are many different choices of similarity one can use at this stage (e.g. percentage of votes in common). We performed our analyses with a common vote similarity as well and found similar results.

This accomplishes an initial dimension reduction from *m* (the number of votes) to $k_0$, the number of clusters in the first scale. We can repeat the above procedure on the residual $R^{(1)} = V - A^{(1)}$ to uncover an additional layer of structure. That is, we can cluster the residual time series and compute the corresponding motivations defined as the means of the new clusters. We can iterate this process and each additional layer gives a next term in what becomes a better and better approximation to the original data. If we have *L* layers with motivations $\{M_1^l, \ldots, M_{k_0(l)}^l\}$ and coefficients $\{\alpha_j^l(i)\}$, (where *l = 1,...,L* and $k_0(l)$ denotes the number of clusters found at scale or level *l*) then at each scale we acquire an additional level of approximation given by

$$A^{(l)} = \begin{pmatrix} \sum_{j=1}^{k_0(l)} \alpha_j^l(1) M_j^l \\ \vdots \\ \sum_{j=1}^{k_0(l)} \alpha_j^l(n) M_j^l \end{pmatrix}$$

We repeat this procedure until the correlation structure of the residual data, $R^{(L)} = V - A^{(1)} - \cdots - A^{(L)}$, is indistinguishable from the correlation structure of a randomly reordered version of the residual data. This gives a "multiscale" or "hierarchical" approximation to the roll call data as

$$V \approx A^{(1)} + A^{(2)} + \cdots + A^{(L)}$$

---

[11] If the $\{M_j\}$ are not linearly independent, this usually signifies a poor choice of the parameters. In such a case, we would simply adjust the parameters.

When applied to the House and Senate, in a number of cases (the 3rd, 4th, 5th, 6th, 7th, 9th, 10th, 11th, 23rd, 57th, 58th, and 77th Senates and the 2nd, 3rd, 18th, 31st, 37th, and 44th Houses) a second layer cannot be computed as there are no demonstrably significant clusters. In other Congresses, we can continue to uncover several more layers of information by this test. In what follows, we choose three examples and discuss the information obtained by investigating one additional layer.

At each scale, the projections onto the motivations accomplish some form of dimension reduction (with dimension generally equal to the number of clusters) of the original data. As a further refinement, these lower-dimensional approximations can be analyzed in and of themselves, for example, via multidimensional scaling (MDS – see Borg and Groenen 2005). Further analysis of the motivations is accomplished via the use of the AdaBoost algorithm (Freund and Schapire 1997). In general, AdaBoost identifies which portions of a data stream best distinguish among a list of clusters. In our case, we can use AdaBoost to identify roll call votes which best determine the boundaries of the clusters of legislators. We interpret the issues behind these votes as the issues that best separate (and hence help define) the motivations associated to the clusters.

It is clear that clustering is a key step in the process. This can be accomplished in many ways, but following the PDM methodology, we choose to use a form of *spectral clustering*. Generally speaking, spectral clustering refers to the determination of clusters in high-dimensional or network data via some use of the eigendata of the *graph Laplacian* (see the Appendix A), a matrix related to the similarity matrix (e.g., distance matrix, in the case of geometric data or the adjacency matrix in the case of a network) of the data. We use the eigendata to determine two parameters: the number of clusters, $k_0$, and $l$, a "natural' dimension for embedding the data network. After determining $k_0$ we then embed the network in $l$ dimensions and use the k-means algorithm (see e.g., Duda et al., 2000) to find the best $k_0$ clusters.

For both of these parameters the second eigenvector (i.e., the eigenvector corresponding to the second smallest eigenvalue) of the graph Laplacian[12] will provide the key information. This is the so-called *Fiedler vector,* which we denote as $v_1$. The Fiedler vector is known to encode the coarse global geometry of the network (see, e.g., Chung 1997). As has been shown in a variety of different settings, the Fiedler vector is effectively the solution to a soft version of various formulations of the problem of finding the optimal decomposition (e.g., minimum cut) of a network into two components (Luxburg 2007, Ng, et al. 2004). The Fiedler vector is indexed by the data or network elements and the clusters can be determined by optimally clustering the entries in the Fiedler vector (see e.g., Luxburg 2007) to determine the best decomposition.

We use a variation in which the number of values around which the Fiedler vector values are clustered is an estimate for the number of clusters. To compute this number we fit the distribution of values in $v_1$ by a Gaussian mixture distribution with the number of components ranging from 2 to 20. We then use the Akaike information criterion (AIC)[13] (Akaike 1974), to measure the closeness of fit. In the case where the minimum AIC value is at least 5% smaller than its nearest competitor, we take the number of components in the fitted distribution with minimum AIC as our choice for $k_0$. If there are multiple AIC values within 5% of the minimum, we take the median value of the number of components of those fitted distributions as $k_0$. This last step is somewhat arbitrary, but it prevents drastic changes in the number of clusters unless truly warranted.

---

[12] The Laplacian is symmetric positive semidefinite and for these roll call examples, has 0 as an eigenvalue with multiplicity one.

[13] We also used another standard criterion, the Bayes information criterion, yielding similar results. The number of estimated clusters was, in general, slightly lower than when using the AIC.

After determining the number of clusters to find in the data, the next step is to pick a dimension for embedding the data. This will be the environment that we use to look for the clusters. This initial dimension reduction is our estimate of $l$, the appropriate dimension for the data. It is determined via an estimate of the number of significant eigenvectors in the graph Laplacian. Here, "significant" means those eigenvectors with nonzero eigenvalues that are smaller than would be expected from a suitable null model. This in turn means that these nonzero eigenvalues are less than a worst case estimate of the eigenvalue for a Fiedler vector of reasonable null model for this data. This builds on the fact that the small eigenvalues and corresponding eigenvectors for the graph Laplacian (like the Fiedler vector) encode the basic clustering information for the network or data (see e.g., Luxburg 2007 for a nice discussion of this). The null model is given by randomized roll call data constructed by randomizing the votes of the legislators for each vote. In other words, for a specific vote, we record the number of yes and no votes and then, for the randomized version of that vote, assign these yes and no votes randomly among the legislators. Executing this over many votes destroys any structure or affinity between groups of legislators.

More precisely, for each set of roll call data, we created 25 randomized versions of the roll call data and computed the Fiedler value for each one. We choose $l$ to be the number of nonzero eigenvalues of the Laplacian associated to the roll call data that are less than the minimum of all Fiedler values for the simulated roll calls. This procedure ensures that the information encoded in the eigenvectors and eigenvalues that we use is distinguishable from "information" that may have arisen due to chance.

We emphasize that this fixed method of parameter selection is a use of unsupervised learning – we do not finesse our choices of parameter or method for individual instances or using additional information. We do this for two reasons. First, we wish to make the method as transparent as possible. Second, we view this method as possibly a first step, subject to refinement in individual cases as warranted. That is, this initial estimate of dimensionality then can be analyzed further (e.g., via multidimensional scaling) in order to achieve a better understanding of the structure in the reduction. This is accomplished in some of our examples. Later on, we will indicate where the results may be pointing towards a refinement in parameter choice and/or method.

We now summarize the method used in the examples below:

1) Compute the correlation matrix $S$ from the data, $V$.

2) Form the graph Laplacian and find its eigenvalues and eigenvectors.

3) Determine $k_0$, the number of clusters, via an AIC-based analysis of the Fielder vector.

4) Determine $l$, the number of significant eigenvectors of the graph Laplacian via comparison with random models.

5) Embed the roll call data in $l$-dimensional Euclidean space via the coordinates given by the first $l$ eigenvectors of the graph Laplacian. Use k-means to find $k_0$ clusters. Determine the motivations as the mean votes of all members of each cluster.

6) Project the data onto the subspace determined by the motivations, forming $A^{(1)}$.

7) Find the residual data $R = V - A^{(1)}$

8) If the residual is not indistinguishable from random noise, repeat steps 1—7 to form the next layer, otherwise stop.

Remarks:

1. While we choose to do at most two iterations of the procedure, it can be iterated until the stopping condition in Step 8 is met.

2. As we will see below, it is useful to apply multidimensional scaling after determining the clusters to get a better idea of the true dimensionality of the first (or depending on the iteration, $l^{th}$) layer.

3. Upon projecting the data onto the motivations, AdaBoost can be (and is) used to determine the votes that best distinguish the clusters.

In the next section we see the process in action.

## Examples

In this section we explore three examples, the $108^{th}$, $88^{th}$, and $77^{th}$ Senates. To provide further intuition for our methodology we begin with a more detailed analysis of the Fiedler vector. Recall that the number of clusters in the entries in the Fiedler vector will serve as our estimate of the number of clusters in the data. This approach is motivated by various analytic and geometric properties of the eigenvector. We try to provide a more intuitive rationale by showing how this simple piece of spectral information actually encodes a great deal of explanatory power for the roll call data. We then continue on to give the full analysis for these data sets.

**The Roll Call Fiedler Vector.** Although it need not be true generally for a network, it turns out that in the case of roll call data, the Fiedler vector $v_1$ is highly localized – most of its values are concentrated near a small list of values. In the Congressional roll call data often the values of $v_1$ (indexed by the legislators) are concentrated around two values, one negative and one positive. Thus, for most Congresses just using the sign of the entries of $v_1$ sorts the legislators into two groups. This generally recovers the basic party split revealed by the NOMINATE models, but other more interesting divisions can also occur.

In Figure 1, we display the results for the $108^{th}$, $88^{th}$, and $77^{th}$ Senates. In each graph the entries of $v_1$ are plotted against the senator indices. These have been listed so that the first block is composed of all the Democrats, the next is all the Republicans, and finally there are the Independents (if any). We now discuss these examples in some detail.

(INSERT FIGURE 1 HERE)

Example 1 – the 108th Senate. Here the values are tightly clustered around 0.1 and -0.1 and proximity to these values essentially distinguishes the two parties. The most "out of place" member is Sen. Miller (D-GA) whose coordinates (11,-0.08474), are marked with a square containing an "x." Sen. Miller is a Democrat who often broke with his party to support the Republican position and who attended and spoke at the 2004 Republican National Convention, endorsing President G. W. Bush for re-election. There are several other members (marked by filled squares) who differ significantly from their closest cluster. Among the Democrats are Senators Baucus, Breaux, Lincoln, and B. Nelson. Among the Republicans are Senators Chaffee, Collins, Snowe, and Spector. These senators are known to break with their party on certain issues.

Example 2 – the 77th Senate. The picture here is much less clear – the Democrats mostly have positive values while the Republicans mostly have negative values. However, there are a significant number of members of both parties with values very close to zero. There are twenty-two senators, marked by filled black squares, with values between -0.05 and 0.05. Moreover, seven Democrats (each marked by an "x"), Senators Adams, Bulow, D. Clark, J. Clark, McCarren, Walsh and Wheeler, lie within the cluster defined by the negative values of the Fiedler vector that is generally associated with the Republican senators. All of these senators broke with the Democratic party on various important issues, indicating a closer ideological alignment with the Republican Party than generally evidenced by a Democrat.

Example 3 – the 88th Senate. Here we have a rather different picture. The last graph in Figure 1 shows the Fiedler vector data for the 88th Senate with Democrats listed first, followed by Republicans with color annotation by region (red = midwest, blue = northeast, green = south, black = southwest, yellow = west).

We see Democrats split into two basic groups of different signs and Republicans, while having mostly negative values, also have a number of members with positive values. We see that in this case, one aspect of the Fiedler vector is associated with region. This is not surprising as one of the signature pieces of legislation in the 88th Congress was the Civil Rights Act of 1964 for which voting split by regional as well as party lines.

This initial investigation of the 108th, 88th, and 77th Senates gives evidence for our claim that the Fiedler vector provides a coarse classification of the Congress in question. Our interpretation of the analysis of the examples is that the Fiedler vector captures party loyalty. In Appendix B, we present simulations that support the claim that the Fiedler vector provides a useful simple classification tool for roll call data.

**The Full PDM Analysis.** The analysis of the Fiedler vector and its classification strength provides one view of the question of dimensionality of the representation which fits well with the results of the NOMINATE models, echoing the finding of one or two dominant dimensions for most of the U.S. Congresses. However, the Fiedler vector is just the first piece of our multidimensional PDM-based analysis.

As we outlined above, a more detailed analysis of the Fiedler vector provides the estimate of $k_0$, the appropriate number of clusters in the data, which are then actually determined in $l$-dimensional space, for a choice of $l$ also guided by the roll call Fiedler vector. In so doing we arrive at an initial dimension reduction, giving an approximation of the originally $m$-dimensional roll call data (where $m$ is the number of votes) as $k_0$-dimensional data determined by the projection onto the space spanned by the motivations. This process is then repeated on each successive residual.

The initial dimension reduction given by the number of clusters ignores the relationships between the clusters. For example, it could be that all the clusters effectively collect on a single line in $l$-dimensional space. We can use multidimensional scaling to provide an estimate of the aggregate dimension of the approximation that takes into account these interrelationships. To do this for the first layer, we compute the matrix of pairwise Euclidean distances between the weighted sums of the motivations for legislators by calculating the length[14] of the difference between the approximation vectors: $d_{i,j} = ||A^{(1)}(i) - A^{(1)}(j)||$. We then compute the multidimensional scaling of this dissimilarity matrix for dimensions one through ten and the stress of each representation. The estimate of the dimension is then the interpolated dimension between one and ten where the stress reaches 0.1. This is a commonly used cutoff point for this method (see e.g. Borg and Gronen, 2005). We repeat the analysis for the second layer (found in the residual as determined by removing the first layer), using $d_{i,j} = ||A^{(2)}(i) - A^{(2)}(j)||$, as well as for the combination of the first and second layer, where $d_{i,j} = ||(A^{(1)}(i) + A^{(2)}(i)) - (A^{(1)}(j) + A^{(2)}(j))||$.

Figure 2 presents the results of doing this MDS analysis on the first two layers for all U.S. Congresses. As indicated, the height of the blue bar is the dimension of the first layer, the height of the red bar is the dimension of the second layer and the black curve gives the dimension of the model produced by combining the two layers. The black curve further gives an indication of the relative strengths of the two layers and the interaction between them.

(INSERT FIGURE 2)

Our results support the basic results obtained by the NOMINATE models – the first layer is most often one-dimensional with notable exceptions such as the period encompassing the 72$^{nd}$ through 90$^{th}$ Senates. However, in the second layer, we see a more complicated picture with many Houses and Senates showing high-dimensional second layers, particularly in periods which precede a jump in the dimension of the first layer or as the dimension of the first layer returns to one.

---

[14] The length a vector $v = (v_1, \ldots v_n)$ in Euclidean space is given by $||v|| = (v_1^2 + \cdots + v_n^2)^{\frac{1}{2}}$.

Our combination of estimates (the black curve) provides a potential explanation of the difference between the NOMINATE and Heckman-Snyder models. Like the Heckman-Snyder factor analysis, we too find many factors/dimensions that contribute to an explanation of the roll call data, but our factors are naturally related to one another in a manner that often produces dimensional estimates in line with the NOMINATE model. This reflects the measurement of scale inherent in our methods – the first layer motivations occur at a coarser scale than those of the second layer and, by the evidence shown in Figure 2, dominate the explanation of most roll call votes. However, the extra dimensions still provide valuable information, creating a high-dimensional "fuzziness" around the low-dimensional approximation, a result that helps explain the necessity of the larger number of dimensions in the Heckman-Snyder factor models.

Carrying through the examples given above, we now give a complete PDM-based analysis of the $108^{th}$, $88^{th}$, and $77^{th}$ Senates. The $77^{th}$ Congress grappled with the entry of The United States into World War II. The hallmark legislation of the $88^{th}$ Congress was the Civil Rights Act of 1964, filibustered for over 14 hours by Sen. Byrd, and splitting the Democratic Senators along regional lines. The $108^{th}$ Congress is considered one of the most politically polarized Congresses since Reconstruction (McCary, et al. 2006), with the vast majority of votes predicted accurately by ideological identification alone.

Example 1 – the $108^{th}$ Senate: Poole and Rosenthal find that the $108^{th}$ Senate is well-described by a single dimension that captures party ideology. Our methods reveal two significant layers and thus, more than one dimension's worth of information.

The information in our first layer is consistent with the picture painted by the one-dimensional summary of Poole and Rosenthal, but provides both quantitative and qualitative descriptions of the contributing ideological factors. The second layer provides a new lens through which we can see organization of the Senate that is issue-based (rather than based on party identification).

At the first scale the Fiedler vector splits the 108th Senate into seven clusters. One of the "Republican" clusters contains a single Democrat (Sen. Miller) but otherwise the clusters are uniform in party: three are Republican (including the barely "mixed" cluster containing Miller) and four are Democrats. Using the $\alpha_j(i)$ values as weights on the mean vote vectors of the clusters (motivations) reduces the roll call votes of this Senate to a combination of seven factors. A reduction of this to two dimensions via multidimensional scaling produces a stress equal to 0.0171, implying that distances in this reduction are relatively true. A two-dimensional view has the further advantage of enabling a visualization of the dimension-reduced data. Figure 3 shows the result with each legislator represented by a shape denoting their cluster membership and color giving their party. The centroids (in these coordinates) of the clusters are shown using a large black unfilled symbol representing the cluster.

(INSERT FIGURE 3 HERE)

Figure 3 reveals that in fact, the data is essentially one-dimensional, but presents a representation of the data that bends the usual "Liberal-Conservative" ideological axis into a roughly U-shaped curve with the most liberal/conservative Senators at the top and the most "moderate" at the bottom. For example, the clusters denoted by squares and downward pointing triangles contain "centrist" senators such as Senators Baucus, Breaux, Nelson, Miller, Chaffee, Collins, Spector, etc. The five other homogeneous clusters represent different factions of their respective parties. This one-dimensional (but curved) picture matches very closely with the ordering given by the NOMINATE models, reflecting the consistency of our methods with the spatial models when the Congress in question is strongly polarized.

To further understand the nature of the sub-ideologies present in the clusters, we use the AdaBoost algorithm to determine which roll call votes best distinguish the clusters. Given the geometry shown in Figure 3, perhaps it is not surprising that most are party-line votes focused on tax cuts, homeland security, state fiscal relief, homeland security, AIDS prevention funding in Africa, etc. In fact, on all of the votes identified, the Republican clusters are monolithic, voting almost 100% in unison. Four votes seem to distinguish the Democratic clusters. The first was an amendment to the tax cut bill to reduce the tax cut proposed by the President Bush to $350 billion. All Democrats voted for this bill except 16% of the cluster denoted by a star and 44% of the cluster denoted by the upward pointing triangle. The second vote, a cloture vote on an amendment concerning energy independence, drew support from all the Democrats except for 8% and 36% of the same two clusters. The last vote, on the passage of the U.S.-Australia free trade agreement, drew widespread support except from 15% of the Republicans in the square cluster, 83% of the Democrats in the right pointing triangle cluster, 25% of Democrats in the star cluster and 29% of the Democrats in the upward triangle cluster.

Upon removing the first layer, we consider the second layer determined by the PDM. We first observe that the residual data for 24 senators cannot be reliably distinguished from the random model – in other words, the roll call votes of these senators may be represented by the first layer approximation with a random perturbation. For the remaining senators, the correlation matrix of the residual data reveals a distinct regional slant: the highest correlations are associated to pairs of senators of the same party and often from the same state. The highest five correlations are between Senators Craig and Crapo (ID), Kerry(MA)/Edwards(NC), Corzine/Lautenberg(NJ), Cantwell/Murray(WA). We note that while Kerry and Edwards are not from the same state, they formed the Democratic Party's presidential ticket in 2004. The second layer has three clusters, two of which are almost exactly evenly split by party and one which is 60%—40% Democrats to Republicans. Thus, the dominant party effect of the first layer has been removed and leaves residual structure that is not obviously related to party.

The AdaBoost algorithm finds five votes that significantly distinguish the three clusters in the second layer. Three votes effectively separated the last two clusters from the first: an amendment to S. 1054, dealing with tax collection contracts, a cloture motion on the motion to recommit S. 1637 (JOBS Act) to committee and an amendment to H.R. 4567 (a homeland security bill) concerning port security grants.

The other two votes effectively separate the second cluster from the other two: a motion to table and amendment to S. 1689 which would provide funds for the Iraq war by suspending a portion of the tax reductions in the highest income bracket and a vote on an amendment to S. Con. Res. 95 to fund medical research, disease control, wellness, tobacco cessation and preventative health efforts via an increase in the tobacco tax. As seen in Figure 2, the multidimensional scaling of this layer has dimension two, indicating that the information in the second layer requires two dimensions to capture it adequately. However, the black line gives the dimension of the combined model that includes both the first and second layer. For the $108^{th}$ Senate, the dimension of the combined model is one, indicating that the first layer is substantially stronger than the second, swamping its influence.

Example 2 – the $88^{th}$ Senate: In their analysis of the $88^{th}$ Senate, Poole and Rosenthal find that the NOMINATE spatial model requires two dimensions to adequately explain the roll call structure. Moreover, they identify those dimensions as associated with party and civil rights. In addition, they analyze collections of votes with very low APRE even after including the second dimension (Poole-Rosenthal 2007, p. 61). The issues they identify include: tax rates, impeachments/investigations, education, ethics and workplace conditions.

Using our methods, the first indication that the $88^{th}$ Senate behaves very differently from other Congresses comes from our investigation of the Fiedler vector, where, as we described above, we see geographic concerns play at least as much of a role as party loyalty. Completing the analysis of this Senate, we find four clusters: the largest of these has mixed party membership, while the remaining three have uniform party membership (two Democratic and one Republican).

(INSERT FIGURE 4 HERE)

Here multidimensional scaling produces a dimension reduction from four to two with an embedding stress equal to 0.0427. The accompanying visualization, shown in Figure 4, reveals a much more complicated situation than that produced by the 108th Congress. Here, the full two dimensions are utilized, revealing a much more complex relationship between the clusters and hence the parties. Figure 4 is coded in three ways. The shape of the marker indicates its cluster (with the heavy black shapes showing the centroids of the clusters) while the color indicates party (red=Republican, blue=Democratic). The color of the outline around the marker indicates region (red=Midwest, blue=northeast, green=south, black=southwest, yellow=west). With this coding we have an interpretation of the two dimensions: party identification and geography. We note that while the vertical direction gives a good correspondence with party identification, the horizontal axis does not correspond to geography since the positions of the geographical clusters are different for different parties. Instead, the horizontal axis roughly captures views on race and civil rights. Those on the right hand side tended to vote against the Civil Rights Act while those on the left hand side vote for the Civil Rights Act. Thus, the first partition of clusters captures a mixture of party identification and standing on the issue of civil rights. Thus, this interpretation is similar to Poole and Rosenthal's analysis.

To see this split in another way we use the AdaBoost algorithm to identify the votes that best separate the clusters. This produces the following list:

- Four amendments to the Civil Rights Act of 1964
- Passage of the Civil Rights Act of 1964
- The Gore Amendment to the Social Security Act, authorizing and funding the creation of Medicare.
- An amendment to the Mass Transportation Act, deleting all funding for mass transit.
- Passage of the Area Redevelopment Act

Each of these votes distinguishes between the two Democratic clusters with the Northern and Southern Democrats on opposite sides of each vote. In general, the Republican cluster votes with the Southern Democratic cluster with two exceptions. First, on the Passage of the Civil Rights Act, 54% of the Republican cluster voted with the Northern Democrats for passage. Second, on the Area Redevelopment Act, 100% of the Northern Democrats and 47% of the Southern Democrats voted for passage while every member of the Republican cluster voted against. Thus, the identified votes confirm the impact of the Civil Rights Act on the roll call network structure as well as identify other issues that contribute significantly.

As with the 108$^{th}$ Senate, removing the first layer reveals a subsidiary geometry that, as measured by correlation strength, has a significant geographic component. In the 88$^{th}$ Senate, the pairs with highest correlations are Hill/Sparkman (AL), Bible/Cannon (NV), Ervin/Jordan (NC), Gruening (AK)/Morse (OR), Bryd/Robertson (VA), Keating/Javits (NY), Keating (NY)/Case (NJ), Aiken/Prouty (VT). Again, as with the 108$^{th}$ Senate, the pairings are mostly between members of the same party. We find three clusters in the residual data that are formed of a mixture of senators from both parties ad using AdaBoost, we find three issue-oriented classifying sets of votes:

- **Agriculture:** Three amendments to H.R. 4997 (to establish a feed grain acreage division program). Two amendments to H.R. 6196, the Administration Farm Bill.

- **Ethics:** An amendment to S. Res. 338 (giving the Rules and Administration Committee the power and responsibility to investigate violations of Senate rules) to create an independent bi-partisan Ethics Committee.

- **Taxes:** An amendment to H.R. 8363 (Revenue Act of 1964) to eliminate preferential tax treatment for profits resulting from stock option plans.

These three sets of issue-oriented votes separate the three clusters: the first and second are divided by the Tax vote, the first and third by two Agriculture votes and the Ethics vote and the second and third by Agriculture votes. Again, we note the commonality of some of these factors with those identified by Poole and Rosenthal.

As shown in Figure 2, the dimension of the second layer, as measured by an effective multidimensional scaling, is between one and two, indicating that at least two dimensions are required to fully capture the geometry of the second layer. Moreover, the multidimensional scaling dimension of the combination of the two layers is also two, indicating (similarly to the 108$^{th}$ Senate), that the first layer dominates the second.

Example 3 – the 77$^{th}$ Senate: Poole and Rosenthal estimate that the 77$^{th}$ Senate is two-dimensional with respect to the NOMINATE spatial model. They arrive at this conclusion by identifying a collection of votes (in this case, related to agriculture) that have low APRE votes in their one-dimensional model, but significantly higher APRE if a second dimension is added.

Using our methods, the 77$^{th}$ Senate also presents another case where one dimension is not sufficient to describe the primary factors. Our initial step in the analysis produces eight clusters for the first scale. We see that five are composed primarily of Democrats, two primarily of Republicans and one (the downward pointing triangle) that is 64% Democratic and 36% Republican. Our dimension estimate for the first layer is 2.67 – in other words, we require three dimensions before multidimensional scaling has stress below 0.1 (the stress of the embedding in three dimensions is 0.0778). Nevertheless, a two-dimensional MDS (See Figure 5) gives some indication of the structure.

(INSERT FIGURE 5)

From this we see that, generally, the parties are distinguished, but that there is significant additional geometric structure. Suspecting that regionalism might also play a role here (as it did in the 88$^{th}$ Senate) we examined these clusters by region as well. While two of the clusters show regional bias – one of mostly southern Democrats and one of northern/Midwestern Republicans – the effect is not nearly as pronounced as in the 88th Senate.

Using the AdaBoost algorithm to identify the votes that best separate the clusters produces a number of relevant votes:

- An amendment to and the passage of H.R. 1776, the Lend-Lease bill.
- An amendment to a resolution concerning appointing a replacement senator from West Virginia.
- Passage of H.R. 4646, a bill to stabilize the U.S. dollar.
- Three amendments to and the passage of H. J. Res 237, a bill modifying the neutrality act.
- Two amendments to H.R. 5990, a bill which set agricultural price controls.
- An amendment to S. J. Res 161 (a bill relating to the stabilization of the cost of living) which would guarantee a farmer's cost of production.

Table 1 shows the percentage of yea votes for these votes, demonstrating some of the ideological distinctions between the clusters.

(INSERT TABLE 1)

As with our previous examples, we see echoes of the Poole-Rosenthal results. However, this example most strikingly demonstrates the kind of additional information that the PDM approach can reveal.

Finally, we note that upon removing the first layer, the residual data exhibits no structure that we can distinguish from random models. Thus, the 77th Senate has no second layer revealed by the PDM.

## Qualitative comparison with other models – predictive ability

We next consider the effectiveness as a predictive model of the dimension-reduced representation given by the first and second layers.

As described in our discussion of methodology, our procedure produces coordinates, $\{\alpha_j(i)\}$, for each legislator, yielding a dimension reduction to a vector space with dimension equal to the number of clusters. Using the first partition for legislator *i*, votes are predicted by the sign of each entry (recall that the entries are indexed by the votes) of in the projection onto the first layer given by $A^{(1)}$. When using the projections onto the first two layers, we predict votes using the sign of the entries of $A^{(1)} + A^{(2)}$.

Comparing these predicted votes with the actual votes gives a measure of accuracy, which we use as a means of comparison with the predictive ability of other models in the literature. Table 2 compares the PDM models with D-NOMINATE models as well as two others, the "minority model" and the "random model."

(INSERT TABLE 2)

The "minority model" is the same baseline model used by Poole and Rosenthal to evaluate the NOMINATE predictions: for each vote, every legislator is assigned a vote equal to the outcome of the roll call (*i.e.,* if the vote passes, all legislators are predicted as voting yes for that vote). The "random model" records the numbers of yeas and nays for a given vote and assigns them to the legislators randomly. In each column, we calculate the percentage of votes correctly predicted from each model as well as the APRE. The statistics from the Poole-Rosenthal model are taken from (Poole and Rosenthal 2007). The column for the random model contains the mean APRE for 10 instances of each random model and the maximum and minimum percent correct. We note that our models significantly outperform the D-NOMINATE model, which is, by this measure, close in accuracy to the random model.

We may also compare the results of the PDM to the Heckman-Snyder models with 1 and 6 factors (Heckman and Snyder 2003, Tables 3a and 3b). As these tables contain only statistics for the 80[th] through 100th Congresses, we did not include them in Table 2. For Senates 80 through 100, the first layer PDM outperforms the 6-factor Heckman-Snyder model in all but the 81st and 82nd Senates. The mean of APRE(PDM)-APRE(Heckman-Snyder) is 0.1731. For Houses 80 through 100, the first layer-PDM outperforms the 6-factor Heckman-Snyder model in all Houses with a mean APRE outperformance of 0.25.

We note that, for individual Congresses, the PDM fares the worst for the 79th House where the APRE is 0.54 and for the 81st Senate where the APRE is 0.43. In general, APRE scores are above 0.8 for all Congresses except those between the 68th and the 90th where the mean APRE is 0.76. In particular, we note that the PDM produces accurate results in periods where the NOMINATE model fails: during the Era of Good Feelings (mean APRE 0.81 for the Houses 14-20 and 0.80 for the same Senates) and in the period before the Civil War (mean APRE 0.82 for Houses 32-35 and 0.76 for the same Senates).

For the U.S. House of Representatives, the number of clusters ranges from 4 to 12 (mean 10.1) while for the Senates, the number ranges from 2 to 15 (mean 5.6). As further comparison with NOMINATE models, we recall that the number of clusters is the number of dimensions of information collected on individual legislators and thus are comparable, in terms of parameters, to a spatial model with the same number of dimensions. Poole and Rosenthal report (Poole and Rosenthal 2007, p. 63-64) that for the 32nd, 85th and 97th Houses the spatial model must have 10-15 dimensions to reach high levels of accuracy. Moreover, the 32nd House, for which the spatial model has a particularly poor fit, reaches only 88% accuracy with ten dimensions. In contrast, the first layer of the PDM yields 91, 93, and 95 percent accuracy for these three houses using 7, 11, and 11 clusters. Beyond the additional efficiency, the use of AdaBoost to identify the clusters ideologically provides verifiable descriptions of the cluster dimensions whereas Poole and Rosenthal report "additional dimensions are largely fitting 'noise' in the data" (Poole and Rosenthal 2007, p. 64, Figure 3.4).

## Conclusion

We have introduced a non-spatial model of roll call voting to produce a geometric decomposition of roll call data. Our main tool is the Partition Decoupling Method (PDM) (Leibon, et al. 2008), which we use to identify and isolate multiple layers of structure that capture significant portions of the functional geometry. The PDM approach identifies clusters in the network that generally represent significant geometric features which, by our model, are associated to motivations that guide legislators' votes. Motivations are centroids in the clusters in the roll call data and can be viewed as a quantification of ideology. Generally, the different layers represent scales of different strengths for the encoded motivations. Our one- and two-layer models, used as predictive models for roll call votes, significantly outperform existing spatial models with respect to standard measures.

To each motivation we associate a voting profile. This serves as a proxy for a description of the aggregate ideology of the members of the cluster. We use the AdaBoost algorithm to provide more detailed information, identifying votes (hence issues) that readily differentiate between the various motivations. Using this method on multiple layers provides a textured description of ideology for each legislator.

Another of our main results is evidence that supports and guides the use of spatial models in this context. Our (non-spatial model) methods yield estimates on the dimensionality of the data, simultaneously reinforcing both the findings of Poole and Rosenthal concerning the low dimensionality of roll call data as well as the higher dimensional requirements of Heckman and Snyder's results. Our estimates show that, on a relatively coarse scale, one or two dimensions almost always is enough to capture the vast majority of the structure of the roll call network. However, in order to understand the network on a finer scale, more dimensions are often necessary. Of particular interest is the observation that the subsidiary dimensions grow just before (and after) changes in the number of primary dimensions. Another observation that warrants further scrutiny is the behavior of these dimensions in recent Congresses, where the primary dimension is one while the number of secondary dimensions is often quite high.

Taken as a whole, we have created an unsupervised statistical learning technique capable of generating a description of the evolving political structure within Congress and more generally, within various working social groups whose behavior can be summarized in roll call-type data. We expect this approach to be of great use in future such analyses, either independently or as a companion to spatial models.

## Appendix A - Spectral Clustering

The phrase "spectral clustering" is used to describe any of a number of techniques for finding clusters or neighborhoods in a collection of data that uses the spectral information (i.e., eigenvectors and eigenvalues) of some matrix associated with the data. Generally speaking, the eigenvalues (their distribution and relative values) give an indication of how many clusters to look for and the eigenvectors are used to create coordinates to construct a (presumably) low-dimensional embedding of the data where the clusters are determined. In this sense, the techniques of principal components analysis (PCA) and multidimensional scaling (MDS) can be viewed as related to spectral clustering. As relates to network data, which includes data that comes with a natural coordinate system (so that the nodes are the points and any pair of edges is connected according to a weight that depends on the distance between them), the matrix that we use is usually some form of the Laplacian of the network, a discrete version of the continuous Laplace operator.

A standard version of spectral clustering, widely used for its flexibility, is due to Ng, Jordan, and Weiss (Ng, et al. 2001). Here is how it would be applied to our correlation matrix *S*. First, construct the graph Laplacian, *L*, associated to *S*:

1. Convert the correlation matrix to the spherical distance: $d = \sin(\cos^{-1}(S)/2)$
2. For a given parameter choice[15] $\sigma$, form $S_1 = e^{-\frac{d^2}{\sigma^2}}$ and remove its diagonal entries.
3. Let *D* be the diagonal matrix of column sums of $S_1$.
4. Then,

$$L = I - D^{-\frac{1}{2}} S_1 D^{-\frac{1}{2}} \tag{15.1}$$

Second, find all eigenvectors $\{v_0, v_1, \ldots, v_{n-1}\}$ associated to eigenvalues $0 \leq \lambda_1 \leq \lambda_2 \leq \cdots \leq \lambda_{n-1}$. Note that in fact *L* is symmetric and positive-semidefinite so that it has *n* nonnegative real eigenvalues of which 0 is one with eigenvector $v_0$ given by any constant vector.

The next step is to determine from this information a "natural" number, of clusters, $k_0$, into which to divide the data. After this is done, we determine a dimensionality *l* for the space in which we look for the clusters. We will use the first *l* eigenvectors to provide coordinates for the data (data point *i* is assigned to the point in *l*-space with coordinates given in order by the $i^{th}$ coordinates of each of the first *l* eigenvectors – much like as is done in PCA). With this embedding, k-means is then used to find the best $k_0$ clusters.

Note that there are two parameters here that need to be determined: the number of clusters, $k_0$, and the dimension *l* of the embedding in which we will apply k-means.

---

[15] In our calculations, we always choose $\sigma = 1$. The choice of this parameter is a choice of scale from which to view the system captured by the roll call votes.

Determination of $k_0$. There are various ways to determine $k_0$. We use the eigenvector for the second largest (i.e., first nonzero) eigenvalue, also called the Fiedler vector. The optimization formulation of the second eigenvalue[16] of $L$ indicates that the solution to a relaxed form of an associated minimization problem related to optimal cluster discrimination is given by a clustering of the entries in the Fiedler vector. As described in the body of the text, after computing the Fiedler vector, we then look for the best approximation as a mixture of Gaussians, the number of which will be the best estimate for $k_0$.

Determination of $l$. Again, the optimization characterization of the eigenvalues of $L$ indicates that the "small" eigenvectors (i.e., eigenvectors for small eigenvalues) dictate the best embedding (see also Ng, et al. 2001). For this, we generate a many instances of matrices that respect various statistics of the original Laplacian and keep the eigenvalues (eigenvectors) of the Laplacian for the roll call data that are smaller than the smallest of the nonzero eigenvalues generated in the random Laplacians.

## Appendix B - Simulation

In this section we present the results of applying our methods to simulated synthetic roll call data both to help justify some of the claims made above but also as a measure of effectiveness of the method. First, we create a simulation of roll call votes that are based on party loyalty alone to demonstrate the effectiveness of the Fiedler vector as a classification tool. We simulated a chamber of Congress with 100 members and 500 votes. Half the members were assigned to one party, the other half to a second party. Each member was given a party loyalty score drawn from a $\beta$-distribution (with $\beta = 1$ and $\alpha$ variable). We constructed a "party line vote" and then a simulated roll call vote as follows. For each member, a random number was drawn and compared to the party loyalty score. If the random number was less than the loyalty score, the vote was cast in accordance to the party line vote. If not, then the vote was cast in opposition. For each simulated roll call, we computed the graph Laplacian and Fiedler vector and computed the correlation between the Fiedler vector and the party loyalty score. We repeated this experiment for 100 trials varying $\alpha$ from 1 to 30 in increment of 0.3.

---

[16] Recall that the $n^{th}$ eigenvalue of a real symmetric matrix $L$ can characterized as the max over all vectors $x$ perpendicular to the space spanned by the first $n-1$ eigenvectors of Rayleigh quotient, $x^t L x / x^t x$. See e.g., (Bau and Trefethen 1997).

This simulation revealed two major results. First, the stronger the party loyalty, the greater the localization of the Fiedler vector (see Figure 6). Second, the correlation between party loyalty and Fiedler vector values was very high, with a mean of 0.9835 and variance 0.0001281 over all trials and members. This helps justify our use of the Fiedler vector as a classification tool. We refer the reader to the original paper (Leibon et al 2008, Supplemental Information) for simulation results concerning the effectiveness of the Partition Decoupling Method.

(INSERT FIGURE 6)

## Bibliography

Akaike, Hirotugu (1974). "A new look at the statistical model identification," *IEEE Transactions on Automatic Control* 19 (6): 716–723.

Bau, David and Trefethen, Lloyd N. (1997). *Numerical Linear Algebra,* Philadelphia: Society for Industrial and Applied Mathematics.

Borg, Ingwer and Patrick Groenen (2005). *Modern Multidimensional Scaling: theory and applications* (2nd ed.), New York: Springer-Verlag.

Cahoon, Lawrence S. (1975). *Locating a Set of Points Using Range Information Only.* Ph.D. Dissertation, Department of Statistics, Carnegie-Mellon University.

Cahoon, Lawrence S., Melvin Hinich, and Peter C. Ordeshook (1978). "A statistical multidimensional scaling method based on the spatial theory of voting." In *Graphical Representation of Multivariate Data,* edited by P.C. Wang. New York: Academic Press.

Chung, Fan (1997). "Spectral Graph Theory," *Regional Conf. Series in Math.,* no. 92, Providence, RI, AMS.

Duda, Richard O., Hart, Peter E., and Stork, David G. (2000), *Pattern Recognition*, (2nd ed.), New York: Wiley-Interscience.

Fiedler, Miroslav (1973). "Algebraic connectivity of graphs," *Czechoslovak Mathematical Journal*: **23** (98):298-305.

Fiedler, Miroslav (1989). "Laplacian of graphs and algebraic connectivity," *Combinatorics and Graph Theory* **25**:57-70.

Clinton, Joshua D, Simon Jackman, and Douglas Rivers (2004), "The statistical analysis of roll call data," *American Political Science Review*, 98: 355-370.

Converse, Philip E. (1964). "The nature of belief systems in mass publics," In *Ideology and Discontent,* edited by David E. Apter. New York: Wiley Press.

Enelow, James and Melvin Hinich (1984) *The Spatial Theory of Voting.* New York: Cambridge University Press.

Freund, Yoav and Robert. E. Schapire (1997). "A decision-theoretic generalization of on-line learning and an application to boosting," *Journal of Computer and System Sciences*, 55(1):119–139.

Freund, Yoav and Robert. E. Schapire (1999). "A short introduction to boosting," *Journal of Japanese Society for Artificial Intelligence,* **14**(5):771-780.

Heckman, James N. and James M. Snyder, Jr (1996). "Linear probability models of the demand for attributes with an empirical application to estimating the preferences of legislators," *RAND Journal of Economics,* 28:S142-S189.

Hinich, Melvin and Michael Munger (1994). *Ideology and the Theory of Political Choice*. Ann Arbor, MI: Univeristy of Michigan Press.

Hinich, Melvin and Michael Munger (1997). *Analytical Politics.* New York: Cambridge University Press.

Hinich, Melvin and Walker Pollard (1981). "A new approach to the spatial theory of electoral competition," *American Journal of Political Science,* 25:323-341.

Leibon, Greg, Scott Pauls, Daniel N. Rockmore, and Robert Savell (2008). "Topological structure in the equities market," *PNAS*, 105:20589-20594.

Luxburg, Ulrike (2007). "A tutorial on spectral clustering," *Statistics and Computing*, 17(4):395—416.

McCarty, Nolan, Keith T. Poole, and Howard Rosenthal (1997), *Income Redistribution and the Realignment of American Politics.* Washington, D.C: AEI Press.

McCarty, Nolan, Keith T. Poole, and Howard Rosenthal (2006), *Polarized America*, New York: MIT Press.

Ng, Andrew Y., Michael I. Jordan, and Yair Weiss (2002), "On spectral clustering: Analysis and an algorithm," In T. Dietterich, S. Becker and Z. Ghahramani, editors, *Advances in Neural Information Processing Systems (NIPS)* 14.

Ordeshook, Peter C. (1976). "The spatial theory of elections: A review and critique." In I. Budge, I. Crewe and S. Farlie, editors, *Party Identification and Beyond.* New York: Wiley

Ordeshook, Peter C. (1986). *Game Theory and Political Theory*. New York: Cambridge University Press.

Poole, Keith T. and Howard Rosenthal (1984). "U.S. presidential elections 1968-1980," *American J. of Political Science,* 28: 282-312.

Poole, Keith T. and Howard Rosenthal (1985). "A spatial model for legislative roll call voting," *American J. of Political Science,* 29: 357-284.

Poole, Keith T. and Howard Rosenthal (2007). *Ideology and Congress*. New York: Oxford University Press.

Poole, Keith T. and Howard Rosenthal (2001), "D-NOMINATE After 10 years: A comparative update to *Congress: A Political-Economic History of Roll Call Voting,*" *Legislative Studies Quarterly*, 26:5-26.

Poole, Keith T. (2005), *Spatial Models of Parliamentary Voting*, Cambridge University Press, Cambridge.

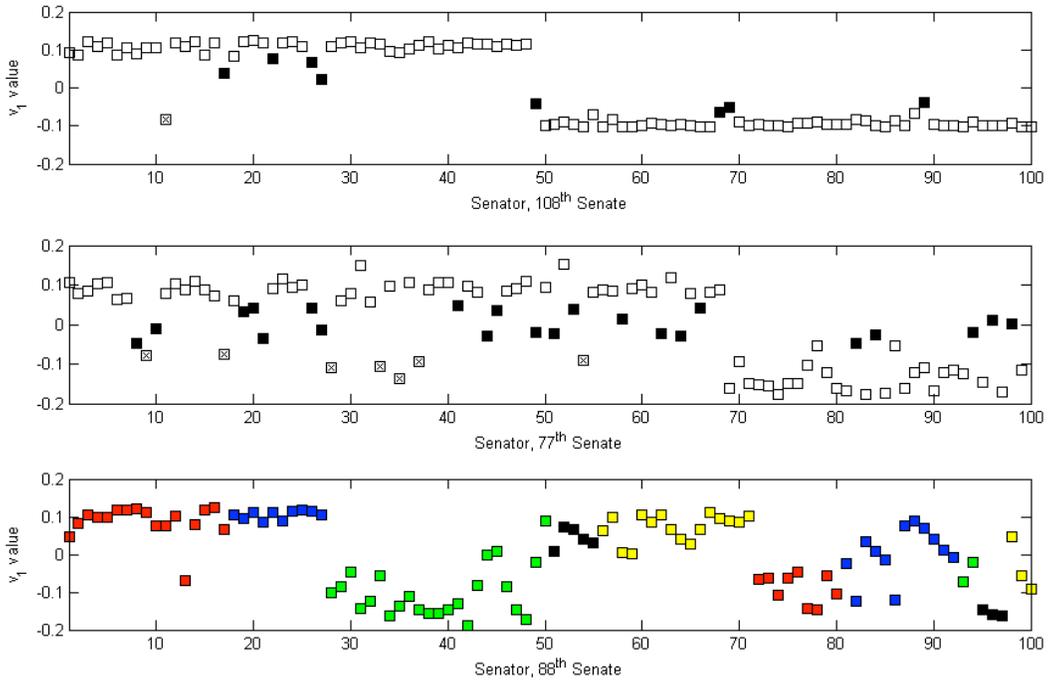

**Figure 1: Fiedler vector values for 108th, 77th, and 88th Senates.** Senators are grouped according to party (Democrats, Republicans and then Independents) in each graph. In the third graph (88th Senate), within the groupings by party, senators are grouped and labeled by region as well (red = Midwest, blue = northeast, green=south, black=southwest, yellow=west).

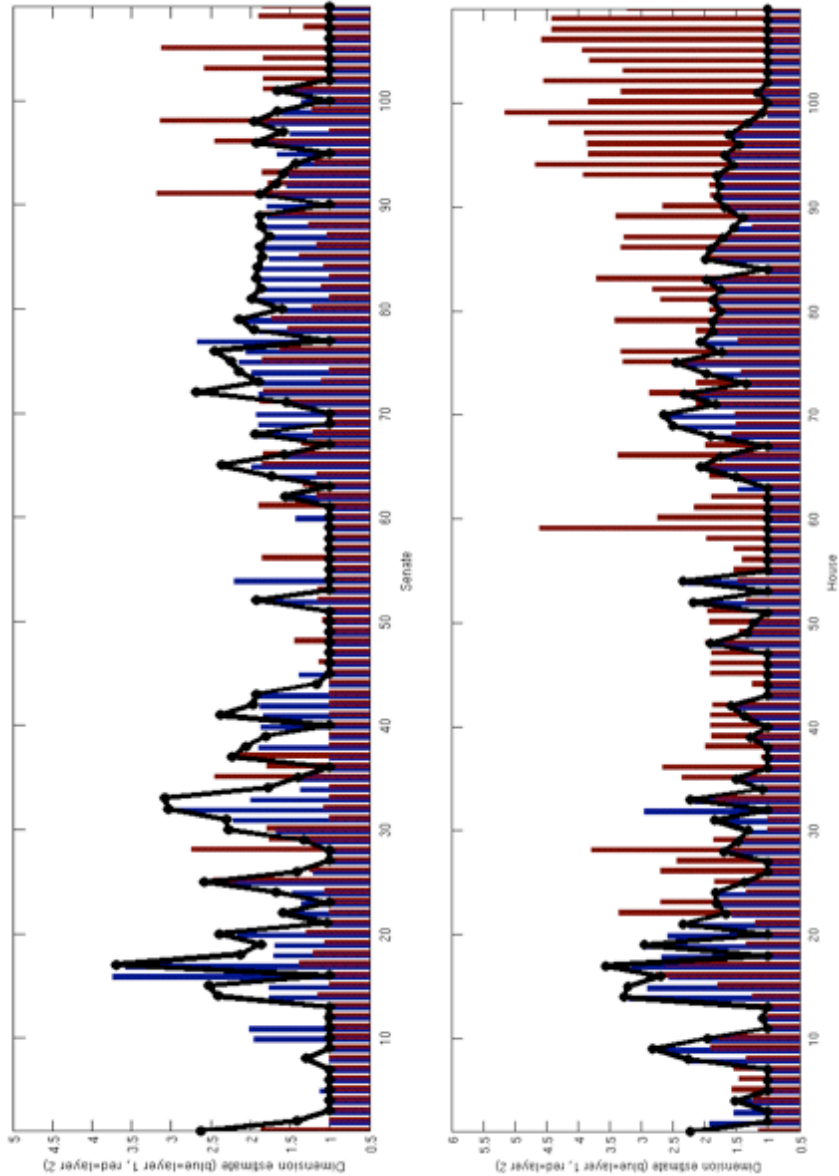

**Figure 2: Dimension estimates via multidimensional scaling.** The height of the blue bar is the MDS estimate of the dimension of the first layer, the height of the red bar is the MDS estimate of the dimension of the second layer and the height of the black curve gives the MDS estimate of the dimension of the model produced by combining the two layers. The height of the black line gives an indication of the relative strengths of the two layers and the interaction between them.

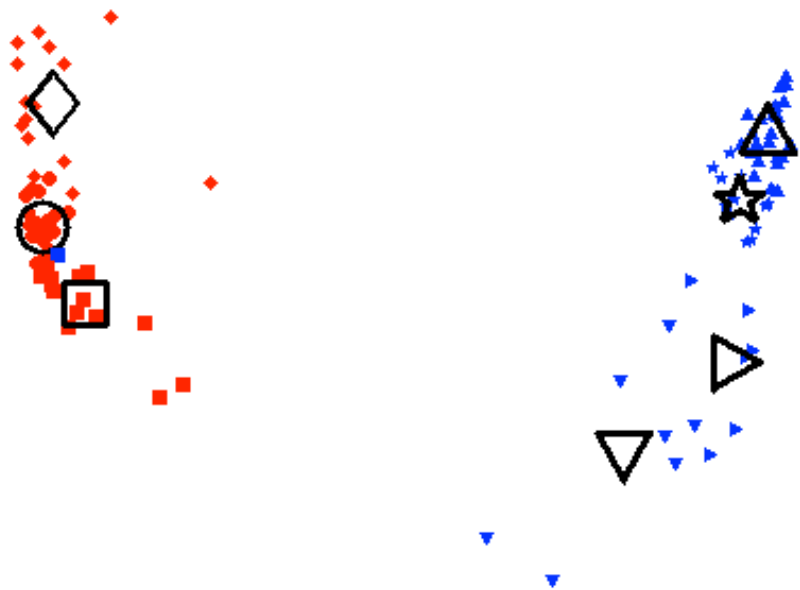

**Figure 3: Two-dimensional MDS representation of the first layer of roll call space for the 108[th] Senate (stress = .01).** Spectral clustering in the first layer determines seven clusters. Cluster membership for individual senators is indicated by one of seven symbols, color-coded red (Republican) and blue (Democrat) with the centroid for the cluster represented by the large black symbol. Note the effective one-dimensionality of the data exemplified by the one-dimensional "U-shaped" structure of the data.

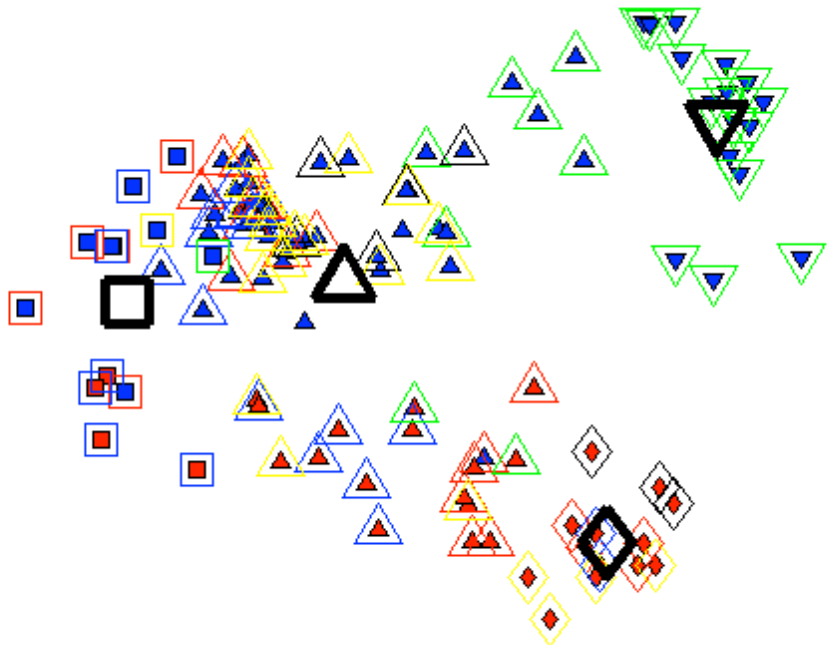

**Figure 4: Two-dimensional MDS representation of the 88th Senate.** Senators are coded in three ways. The shape of the marker indicates its cluster (with the heavy black shapes showing the centroids of the clusters) while the color indicates party (red=Republican, blue=Democratic). The color of the outline around the marker indicates region (red=Midwest, blue=northeast, green=south, black=southwest, yellow=west).

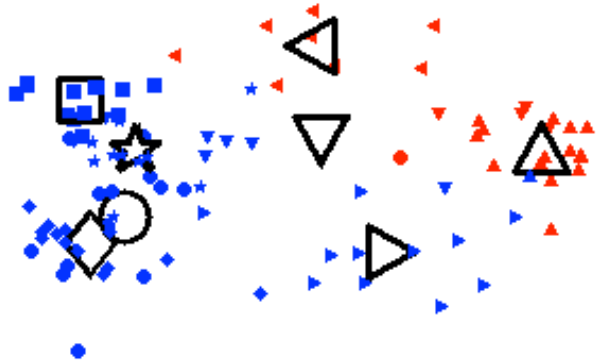

**Figure 5: Representation of the 77$^{th}$ Senate using multidimensional scaling.** Our initial step in the analysis produces eight clusters for the first scale. An "acceptable" MDS requires three dimensions (the stress of the embedding in three dimensions is 0.0778) but the two-dimensional MDS gives an indication the structure (with stress 0.1444). Five clusters are composed primarily of Democrats, two primarily of Republicans and one (the downward pointing triangle) is 64% Democratic and 36% Republican.

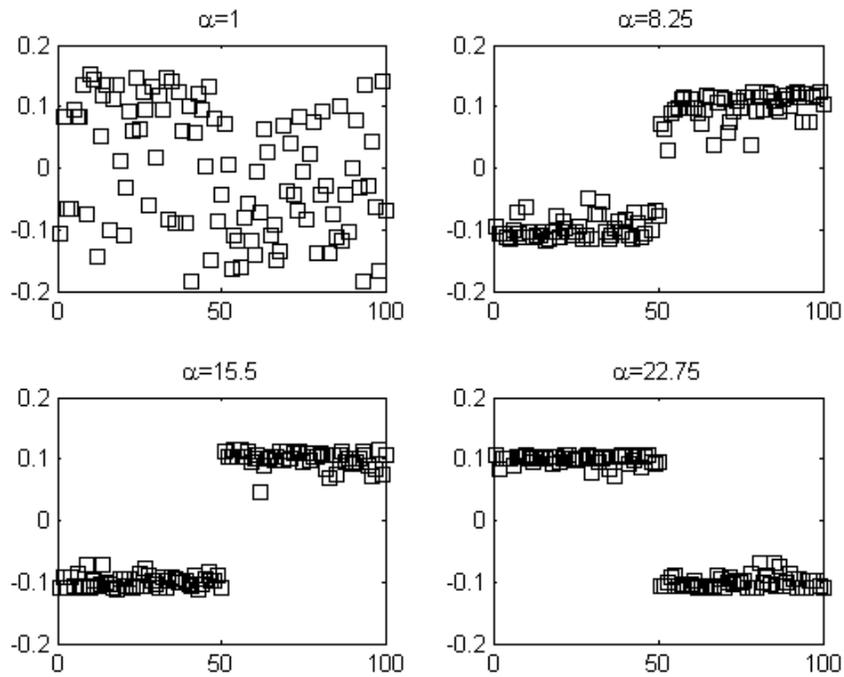

**Figure 6. Fiedler vector plots for four different roll call simulations with differing party loyalty assumptions.** Each plot is labeled by an ♋ value which determines the ♌-distribution (with ♌=1) which models the party loyalty scores for simulated legislators. The plots are the values of the Fiedler vector for a correlation matrix derived from 500 simulated votes.

| Roll call | ▲ | ◀ | ▼ | ★ | ● | ■ | ♦ | ▶ |
|---|---|---|---|---|---|---|---|---|
| 10 | 100 | 100 | 63 | 0 | 0 | 0 | 0 | 82 |
| 19 | 6 | 100 | 55 | 100 | 100 | 100 | 100 | 0 |
| 29 | 82 | 100 | 100 | 0 | 15 | 8 | 20 | 73 |
| 53 | 6 | 0 | 36 | 100 | 100 | 100 | 100 | 89 |
| 80 | 100 | 38 | 64 | 0 | 25 | 0 | 8 | 100 |
| 81 | 100 | 38 | 73 | 0 | 18 | 0 | 8 | 100 |
| 83 | 0 | 63 | 27 | 100 | 83 | 100 | 92 | 0 |
| 84 | 0 | 63 | 27 | 100 | 83 | 100 | 92 | 0 |
| 99 | 0 | 100 | 100 | 20 | 8 | 100 | 15 | 56 |

**Table 1. Votes that best distinguish between clusters in the 77$^{th}$ Senate (as determined by AdaBoost).** Entries of the table are the percent of members in each cluster (column) which vote "yea" on a given roll call vote (row). The clusters are identified by the same symbol used in Figure 5. The clusters are ordered by party from left to right: the two clusters listed to the left are comprised primarily of Republicans, the third cluster is a mixtures of Democrats and Republicans and the remaining clusters are primarily Democrats.

|  | Minority model | Random model | NOMINATE: 1 dim. | NOMINATE: 2 dim. | PDM: one layer | PDM: two layer |
|---|---|---|---|---|---|---|
| House APRE | 0 | 0.4561 | 0.534 | 0.593 | 0.839 | 0.856 |
| Percent correct (House) | 67.3 | [72,88] | 84.5 | 86.5 | 94.7 | 95.3 |
| Senate APRE | 0 | 0.4834 | 0.476 | 0.563 | 0.809 | 0.822 |
| Percent correct (Senate) | 66.6 | [70,90] | 82.3 | 85.2 | 93.6 | 94.1 |

**Table 2. Comparative predictive performance of different models.** Four basic models are compared: the minority model, the random model, the NOMINATE spatial models and the PDM models. Each is evaluated via APRE and percent correct prediction for all U.S. Houses and all U.S. Senates. The minority model is used to construct the APRE while the random model is used as a null model in the PDM. Due to the nature of the random model, the percent correct statistic cannot be computed in aggregate. Instead we report the interval containing all instances (see text).